# Active self-adaptive metamaterial plates for flexural wave control

Zheng-Yang Li[a], Tian-Xue Ma[a, *], Yan-Zheng Wang[a], Yu-Yang Chai[b], Chuanzeng Zhang[a, *], Feng-Ming Li[c]

[a] Department of Civil Engineering, University of Siegen, D-57076 Siegen, Germany

[b] Department of Mechanical Engineering, The Hong Kong Polytechnic University, 999077, Hong Kong, China

[c] College of Aerospace and Civil Engineering, Harbin Engineering University, 150001 Harbin, China

[*] Corresponding author, tianxue.ma@uni-siegen.de, c.zhang@uni-siegen.de

**Abstract**

In this paper, a novel design concept for active self-adaptive metamaterial (ASAMM) plates is proposed based on an active self-adaptive (ASA) control strategy guided by the particle swarm optimization (PSO) technique. The ASAMM plates consist of an elastic base plate and two periodic arrays of piezoelectric patches. The periodic piezoelectric patches place on the bottom plate surface act as sensors, while the other ones attached on the top plate surfaces act as actuators. A simplified plate model is established by the Hamilton principle. By assuming a uniform or constant plate thickness, the plane wave expansion (PWE) method is adopted to calculate the band structures. The finite element method (FEM) using 2D plate and 3D solid elements is also used to calculate the band structures and the transmission spectra or frequency responses. The conventional displacement, velocity and acceleration feedback control methods are introduced and analyzed. Then, a novel ASA control strategy based on combining the displacement and acceleration feedback control methods and guided by the PSO technique is developed. Numerical results will be presented and discussed to show that the proposed ASAMM plates can automatically and intelligently evolve different feedback control schemes to adapt to different stimulations on demand. Compared to the conventional metamaterial (MM) plates, the proposed ASAMM plates exhibit improved and enhanced band-gap characteristics and suppression performance for flexural waves at frequencies outside the band-gaps.

## 1. Introduction

Elastic or acoustic metamaterials (MMs) are artificially periodic or non-periodic structures, which may show unprecedented properties that barely or even not exist in nature, such as negative effective stiffness, negative effective mass density, negative refraction, band-gaps, etc. These unique characteristics have attracted a great deal of attention in recent years to manipulate the elastic/acoustic wave propagation properties in the elastic/acoustic MMs [1, 2]. A variety of unconventional wave manipulation measures have been proposed over the last decades, such as the wave front manipulation [3], flexural wave steering [4], flexural wave black hole [5], and so on. However, the conventional elastic/acoustic MMs often suffer from the rather narrow frequency bandwidth, which restricts their further developments and applications. Hence, tunable MMs (TMMs) have been proposed to overcome this limitation of the conventional elastic/acoustic MMs.

The TMMs can manipulate the elastic/acoustic wave propagation flexibly and thus enhance the efficiency and performance of the conventional MMs. The tunning strategies of the TMMs can be roughly classified into two major categories. The first one is the topological reconfiguration, such as using the buckling of the soft porous materials to tune the band-gaps [6]. The temperature is applied to the MM to adjust the effective mass density through the topological changes induced by the thermal buckling modes [7]. The other one is combined with the tunable materials, such as using different water levels to tune the dispersions of evanescent waves [8]. The helical MM manufactured with the three-dimensional (3D) printed cylindrical units capable of tuning acoustic waves in a wide frequency range by altering the insertion depth of the helical cylinders was investigated [9]. Moreover, piezoelectric materials are frequently used as the smart or intelligent tunable material by utilizing their electromechanical coupling effect and quick response ability.

As a typical and representative piezoelectric material, the lead zirconate titanate (PZT) has been widely used as the smart or intelligent material for noise and vibration control in the last decades. Shunted PZT patches were used to control the longitudinal wave propagation in rods [10]. The control of flexural waves in the elastic beams was realized by a periodic array of resistive inductive shunted PZT patches both theoretically and experimentally [11]. Furthermore, the control of the flexural elastic waves by shunted PZT patches of different circuits in the elastic plates was also proposed [12, 13]. Based on the idea of active control, the active feedback control technique using shunted PZT patches to manipulate the flexural waves in the elastic beams was developed [14-16]. Recently, a lightweight and adaptive hybrid laminate MM based on PZT patches was designed and investigated [17].

Based on the studies of noise and vibration control, the PZT patches with shunted circuits were previously used to build the TMMs. The TMM consisting of shunted PZT patches was suggested to realize an extremely broadband control of flexural waves in the elastic beams by the local resonance mechanism [18]. The adaptive TMM was proposed [19]. A tunable beam-type TMM composed of elastic base beams and periodically arrayed PZT patches was designed to achieve a low-frequency band-gap [20]. The focusing lens was built by the PZT patches [21]. Furthermore, a broadband controllable stiffness TMM was presented to realize non-reciprocal wave propagation and tunable band-gaps [22]. The TMM beams with a digitally tunable band-gap [23] and the local resonant piezoelectric MM plates [24] were investigated.

More recently, the self-actuating and self-sensing materials were proposed inspired by the natural creatures altering their functionality in response to various external stimulations. We refer to Ref. [25] for a detailed review in these research fields. Base on the similar idea, scholars have achieved remarkable progresses in recent decades [2]. The intelligence to be programmable to dynamically and arbitrarily manipulate electromagnetic wave fields was conveyed to materials [26]. The self-adaptive wave cloak driven by deep learning was studied theoretically and experimentally [27]. However, to the best of our knowledge, the abovementioned control measures are rarely adopted and applied for controlling the elastic and acoustic waves [28].

In this work, active self-adaptive metamaterial (ASAMM) plates consisting of an elastic base plate and two periodic arrays of piezoelectric patches for elastic flexural wave or vibration manipulation as shown in Figs. 1(a)-(c) are proposed and investigated. The period piezoelectric patches on the bottom plate surface act as the sensors while the other ones on the top plate surface act as the actuators. Based on the active displacement and acceleration feedback control schemes, an active self-adaptive (ASA) control strategy is established by adopting the particle swarm optimization (PSO) technique [29, 30] to realize the self-sensing and self-actuating of the sensors and actuator for suppressing the elastic flexural waves. We refer to Ref. [31] for a detailed review on the PSO technique. Hence, the designed ASAMM plates in this paper can automatically sense the external stimulation and act correspondingly guided by the proposed ASA control strategy. Unlike many previous studies related to the present investigation, which were mostly focused only on the widths and locations of the band-gaps, the novelty of the present work lies in the fact that it will tackle the following three issues simultaneously (see Fig. 1(d)):

- **Enhancement of the band-gap characteristics:** This includes the band-gap widening and the attenuation increase within the band-gaps.
- **Flexural wave or vibration suppression at frequencies outside the band-gaps:** The resonance peaks in the transmission or response spectra outside the band-gaps should be drastically reduced.
- **Self-adaptivity:** In particular, the designed ASAMM plates can perform the abovementioned two tasks automatically without human intervention.

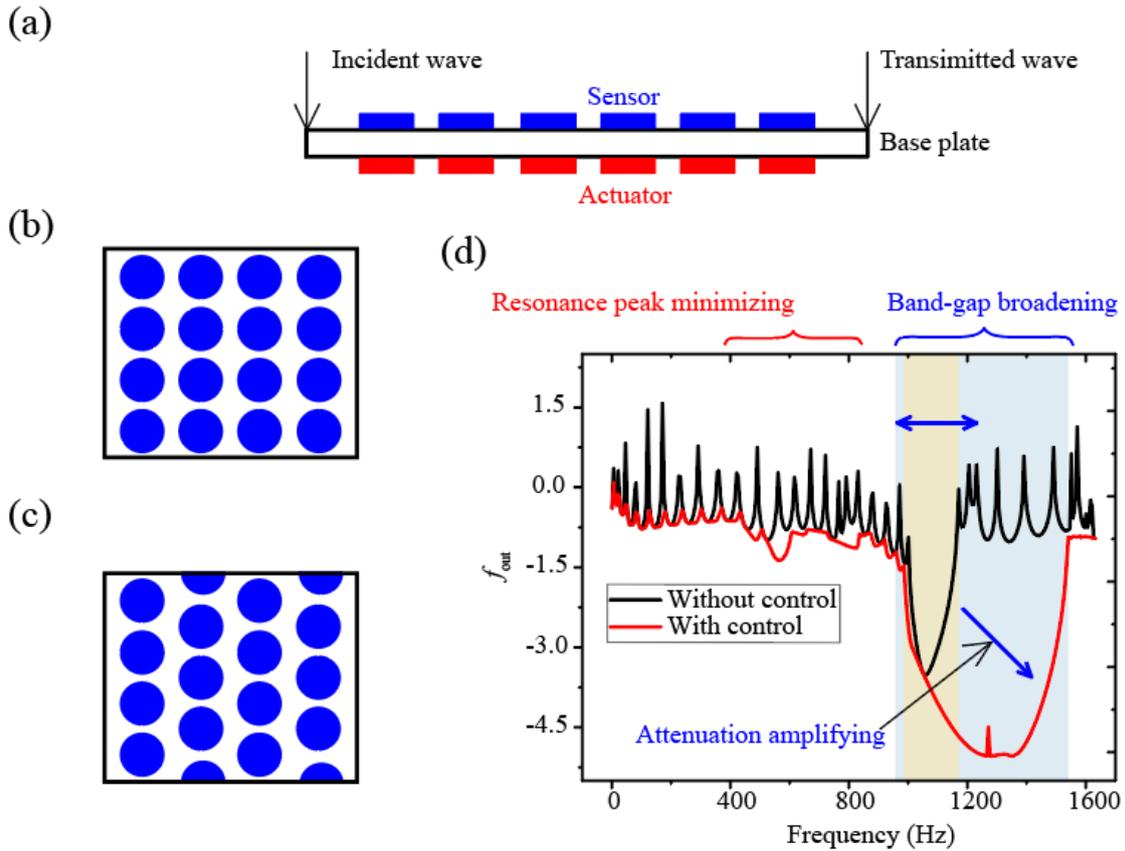

Fig. 1: Active self-adaptive metamaterial plate (a) with square-latticed (b) and triangle-latticed (c) piezoelectric patches as sensors and actuators. An example for the transmission spectra is illustrated in (d) to highlight the key objectives of the present study. The shaded areas in (d) mark the band-gaps without (yellow) and with (gray) the ASA control strategy.

The designed ASAMM plates can further enhance the vibration isolation performance of the lattice sandwich structures [32], and the proposed ASA control strategy can be extended and applied to design novel elastic/acoustic metasurfaces, wave clocks, lenses and other elastic/acoustic devices on demand.

## 2. Theoretical modeling, band structure and transmission calculations

### 2.1 Description of the MM plate model

Figure 2 shows the unit-cell of the MM plate with piezoelectric patches as the actuator/sensor, which are periodically placed at the upper/lower surface of the base plate with the lattice constant $a$. The thickness of the base plate is $h_b$, where the subscript "b" stands for the base plate. The radius and thickness of the piezoelectric patches are $r_{pa}=r_{ps}$ and $h_{pa}=h_{ps}$, where the subscripts "pa" and "ps" represent the actuator and sensor, respectively. The commercial finite element method (FEM) software COMSOL Multiphysics can be used to calculate the band structures and frequency responses based on 3D or volume elements without simplifications, but it is inefficient computationally expensive. Therefore, a simplified model based on the Kirchhoff thin plate theory is developed in this work to calculate the band structures and frequency responses efficiently and accurately, which is described in the following.

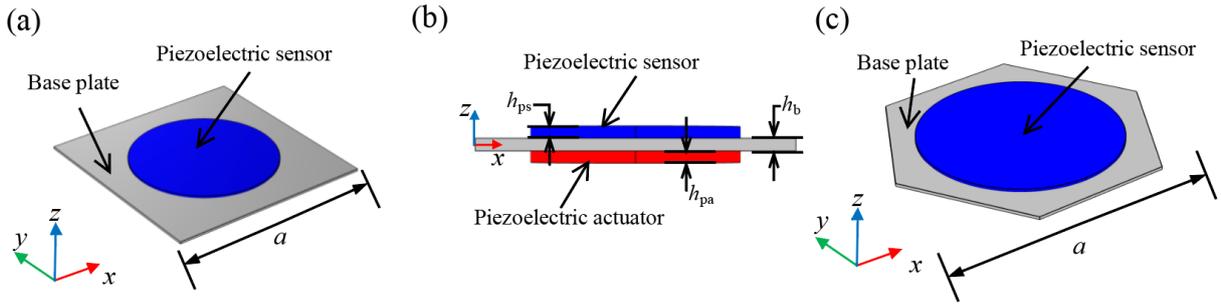

Fig. 2 Unit-cells of the MM plates in the square lattice (a, b) and the triangle lattice (c).

The base plate is a continuous linear elastic thin plate, and thus the Kirchhoff thin plate theory can be applied. The stresses and strains of the plate are given by

$$\boldsymbol{\sigma} = \left[\sigma_x, \sigma_y, \sigma_{xy}\right]^{\mathrm{T}} = \left[-\frac{Ez}{1-\nu^2}\left(\frac{\partial^2 w}{\partial x^2}+\mu\frac{\partial^2 w}{\partial y^2}\right), -\frac{Ez}{1-\nu^2}\left(\frac{\partial^2 w}{\partial y^2}+\mu\frac{\partial^2 w}{\partial x^2}\right), -\frac{Ez}{1+\nu}\frac{\partial^2 w}{\partial x \partial y}\right]^{\mathrm{T}}, \quad (1)$$

$$\boldsymbol{\varepsilon} = \left[\varepsilon_x, \varepsilon_y, \varepsilon_{xy}\right]^{\mathrm{T}} = \left[-z\frac{\partial^2 w}{\partial x^2}, -z\frac{\partial^2 w}{\partial y^2}, -2z\frac{\partial^2 w}{\partial x \partial y}\right]^{\mathrm{T}}, \quad (2)$$

where $w$ is the transverse displacement of the plate, $E$ is the Young's modulus, $\nu$ is the Poisson's ratio, $\sigma_x$ and $\sigma_y$ are the normal stresses, $\sigma_{xy}$ is the in-plane shear stress, $\varepsilon_x$ and $\varepsilon_y$ are the normal strains, $\varepsilon_{xy}$ is the in-plane shear strain, and $z$ is the transverse coordinate perpendicular to the plane of the plate.

The equation of motion of the base plate without the piezoelectric patches can be established as [15]

$$D_b^E \nabla^2 \nabla^2 w = m_b \frac{\partial^2 w}{\partial t^2}, \qquad (3)$$

in which

$$\nabla^2 = \frac{\partial^2}{\partial x^2} + \frac{\partial^2}{\partial y^2}, \quad D_b^E = \frac{E_b h_b^3}{12(1-v_b^2)}, \quad m_b = \rho_b h_b, \qquad (4)$$

where $E_b$ and $v_b$ are the Young's modulus and Poisson's ratio of the base plate.

Under the assumption of plane stress for a thin plate, the reduced constitutive equations of the piezoelectric patches can be expressed as (33)

$$\begin{bmatrix} \varepsilon_x \\ \varepsilon_y \\ \varepsilon_{xy} \\ D_z \end{bmatrix} = \begin{bmatrix} s_{11}^E & s_{12}^E & 0 & d_{31} \\ s_{21}^E & s_{22}^E & 0 & d_{32} \\ 0 & 0 & s_{66}^E & 0 \\ d_{31} & d_{32} & 0 & \varepsilon_{33}^T \end{bmatrix} \begin{bmatrix} \sigma_x \\ \sigma_y \\ \sigma_{xy} \\ E_z \end{bmatrix} \quad \text{or} \quad \begin{bmatrix} \sigma_x \\ \sigma_y \\ \sigma_{xy} \\ E_z \end{bmatrix} = \begin{bmatrix} c_{11}^E & c_{12}^E & 0 & -e_{31} \\ c_{21}^E & c_{22}^E & 0 & -e_{32} \\ 0 & 0 & c_{66}^E & 0 \\ e_{31} & e_{32} & 0 & \varepsilon_{33}^S \end{bmatrix} \begin{bmatrix} \varepsilon_x \\ \varepsilon_y \\ \varepsilon_{xy} \\ D_z \end{bmatrix} \qquad (5)$$

where $s_{11}^E$, $s_{12}^E$, $s_{21}^E$ and $s_{66}^E$ are the elastic compliance constants under constant electric field, $d_{31}$ and $d_{32}$ are the piezoelectric constants, $\varepsilon_{33}^T$ is the dielectric permittivity under constant stress, $c_{11}^E$, $c_{12}^E$, $c_{21}^E$ and $c_{66}^E$ are the reduced elastic constants, $e_{31}$ and $e_{32}$ are the reduced piezoelectric constants, $\varepsilon_{33}^S$ is the reduced dielectric permittivity under constant strain, $D_z$ is the electric displacement, and $E_z$ is the electric field. For transversely isotropic piezoelectric materials, the reduced elastic, piezoelectric and dielectric constants are determined by [33]

$$c_{11}^E = c_{22}^E = \frac{s_{11}^E}{(s_{11}^E + s_{12}^E)(s_{11}^E - s_{12}^E)}, \quad c_{12}^E = \frac{-s_{12}^E}{(s_{11}^E + s_{12}^E)(s_{11}^E - s_{12}^E)}, \quad c_{66}^E = \frac{1}{s_{66}^E},$$

$$e_{31} = e_{32} = \frac{d_{31}}{s_{11}^E + s_{12}^E}, \quad \varepsilon_{33}^S = \varepsilon_{33}^T - \frac{2d_{31}^2}{s_{11}^E + s_{12}^E}.$$

Throughout the analysis, it is assumed that the polarization direction of the piezoelectric patches is along the $z$-direction.

For the part of the base plate covered by the piezoelectric patches, the following Hamilton principle can be used to establish the electromechanically coupled equations

$$\int_{t_1}^{t_2} \delta(T-Q)dt + \int_{t_1}^{t_2} \delta W dt = 0, \qquad (6)$$

where $\delta$ is the first variation, $T$ is the kinetic energy, $Q$ is the potential energy, and $\delta W$ is the virtual work of the externally applied forces, which can be ignored for the wave propagation or free vibration analysis. The kinetic energy of the composite plate with the piezoelectric patches is given by

$$T = \left(\frac{\rho_b h_b}{2} + \frac{\rho_{pa} h_{pa}}{2} + \frac{\rho_{ps} h_{ps}}{2}\right) \int_S \left(\frac{\partial w}{\partial t}\right)^2 dS = m_p \int_S \left(\frac{\partial w}{\partial t}\right)^2 dS, \qquad (7)$$

where $S$ is the cross-section area of the PZT and base plate, and $m_p$ is the mass per area of the cross-sectional area defined by

$$m_p = \frac{\rho_b h_b}{2} + \frac{\rho_{pa} h_{pa}}{2} + \frac{\rho_{ps} h_{ps}}{2}.$$

The potential energy of the composite plate with the piezoelectric patches is

$$Q = \frac{1}{2}\int_{V_b} \boldsymbol{\varepsilon}^T \boldsymbol{\sigma} dV_b + \frac{1}{2}\int_{V_{pa}} \boldsymbol{\varepsilon}^T \boldsymbol{\sigma} dV_{pa} - \frac{1}{2}\int_{V_{pa}} D_{zpa} E_{zpa} dV_{pa} + \frac{1}{2}\int_{V_{ps}} \boldsymbol{\varepsilon}^T \boldsymbol{\sigma} dV_{ps} - \frac{1}{2}\int_{V_{ps}} D_{zps} E_{zps} dV_{ps}, \qquad (8)$$

where $D_z$ is the electric displacement of the piezoelectric patches, $E_z$ is the electric filed intensity of the piezoelectric patches, and $V$ is the volume. Equation (8) can be divided into three parts, the potential energies from the base plate, the piezoelectric sensor, and the piezoelectric actuator. The contribution from the base plate is

$$Q_b = \frac{1}{2}\int_{V_b} \boldsymbol{\varepsilon}^T \boldsymbol{\sigma} dV_b = \frac{1}{2}\int_{V_b} \frac{E_b z^2}{1-v_b^2}\left(\nabla^2 w\right)^2 dV_b = \frac{1}{2}D_b^E \int_S \left(\nabla^2 w\right)^2 dS. \qquad (9)$$

For the piezoelectric actuator, we have

$$Q_{pa} = \frac{1}{2}\int_{V_{pa}} \boldsymbol{\varepsilon}^T \boldsymbol{\sigma} dV_{pa} - \frac{1}{2}\int_{V_{pa}} D_{zpa} E_{zpa} dV_{pa}$$

$$= \frac{1}{2}D_{pa}^E \int_S \left(\nabla^2 w\right)^2 dS + \frac{1}{2}C_{pa}^P \int_{V_{pa}}\left(e_{31} E_{zpa} \frac{\partial^2 w}{\partial x^2} + e_{31} E_{zpa} \frac{\partial^2 w}{\partial y^2}\right) dV_{pa} + \frac{1}{2}\int_{V_{pa}} \varepsilon_{33}^S E_{zpa}^2 dV_{pa}, \qquad (10)$$

and

$$D_{pa}^{E} = \frac{E_{pa}h_{pa}^{3}}{12(1-v_{pa}^{2})}, \tag{11}$$

$$C_{pa}^{P} = \frac{1}{2}\left[\left(\frac{h_{b}}{2}+h_{pa}\right)^{2}-\frac{h_{b}^{2}}{4}\right]. \tag{12}$$

$D_{pa}^{E}$ in Eq. (11) is the flexural rigidity of the piezoelectric actuator. $C_{pa}^{P}$ in Eq. (12) can be considered as the additional flexural rigidity coefficient contributed by the electric field. Since only the low-frequency range of the flexural waves is of particular interest in this analysis, we assume that the electric field in the piezoelectric patches is uniform. Then we have

$$E_{zpa} = \frac{U_{zpa}}{h_{pa}}, \quad E_{zps} = \frac{U_{zps}}{h_{ps}}, \tag{13}$$

where $U_z$ is the out-of-plane voltage of the piezoelectric patch. Substituting Eq. (13) into Eq. (10), we can obtain the contribution to the potential energy from the piezoelectric actuator as

$$Q_{pa} = \frac{1}{2}D_{pa}^{E}\int_{S}(\nabla^{2}w)^{2}dS + \frac{1}{2}C_{pa}^{P}\int_{V_{pa}}\frac{e_{31}}{h_{p}}U_{zpa}\nabla^{2}wdV_{pa} + \frac{1}{2}\int_{V_{pa}}\frac{\varepsilon_{33}^{S}}{h_{p}}U_{zpa}^{2}dV_{pa}. \tag{14}$$

The contribution to the potential energy from the piezoelectric sensor can be obtained in the same form as Eq. (14), but the subscript "a" in Eq. (14) should be replaced by "s". Substituting Eqs. (7)-(14) into Eq. (6), one can obtain the following governing equations

$$m_{p}\frac{\partial^{2}w}{\partial t^{2}}+(D_{b}^{E}+D_{pa}^{E}+D_{ps}^{E})\nabla^{2}\nabla^{2}w+\frac{e_{31}}{h_{p}}C_{pa}^{P}\nabla^{2}U_{zpa}-\frac{e_{31}}{h_{p}}C_{ps}^{P}\nabla^{2}U_{zps}=0, \tag{15}$$

$$e_{31}C_{pa}^{P}\nabla^{2}w - \varepsilon_{33}^{S}U_{zpa} = 0, \tag{16}$$

$$e_{31}C_{ps}^{P}\nabla^{2}w + \varepsilon_{33}^{S}U_{zps} = 0, \tag{17}$$

where $D_{pa}^{E} = D_{ps}^{E}$ and $C_{pa}^{P} = C_{ps}^{P}$. Substituting Eqs. (16) and (17) into Eq. (15), one can derive the open-circuit governing equation for the MM plate as

$$m_{p}\frac{\partial^{2}w}{\partial t^{2}}+D_{p}\nabla^{2}\nabla^{2}w=0, \tag{18}$$

in which $D_p = D_b^E + D_{pa}^E + \left(e_{31}C_{pa}^P\right)^2 / \varepsilon_{33}^S h_p + D_{ps}^E + \left(e_{31}C_{ps}^P\right)^2 / \varepsilon_{33}^S h_p$ is the effective flexural rigidity.

## 2.2 Band structure calculations

An important issue for studying the dynamic characteristics of the considered MM plates is the computation of the frequency band structures or dispersion relations. In this subsection, numerical methods based on the plane wave expansion methods and the FEM will be presented for this purpose. To simplify the analysis, we first assume that the MM plate has a uniform thickness. Therefore, the interfaces of the plate can be neglected. And the MM plate is considered as a two-phase phononic crystal plate [34] as shown in Fig. 3(a). Then we can calculate the band structure by the PWE method [34-36]. By expanding the field quantities into the Fourier series components along the reciprocal lattice vector in Eq. (18), the resulting linear sytsem of algebraic equations can be written as

$$\bar{m}\omega^2 \sum_{\mathbf{G'}} w_{\mathbf{k+G'}}(\mathbf{G'}) = \bar{D}\sum_{\mathbf{G'}} \left(\left(\mathbf{k+G'}\right)_x^2 + \left(\mathbf{k+G'}\right)_y^2\right)^2 w_{\mathbf{k+G'}}(\mathbf{G'}), \quad (19)$$

in which $\mathbf{k}$ is the Bloch wave vector in the first irreducible Brillion zone, $\bar{m}$ and $\bar{D}$ are the Fourier coefficients of $\bar{m}_p$ and $\bar{D}_p$, and $\mathbf{G'}$ is given by

$$\mathbf{G'} = m\mathbf{b}_1 + n\mathbf{b}_2, \quad (20)$$

where $m$ and $n$ are integers, $\mathbf{b}_1 = (b_{1x}, b_{1y})$ and $\mathbf{b}_2 = (b_{2x}, b_{2y})$ are the basis vectors of the reciprocal lattice.

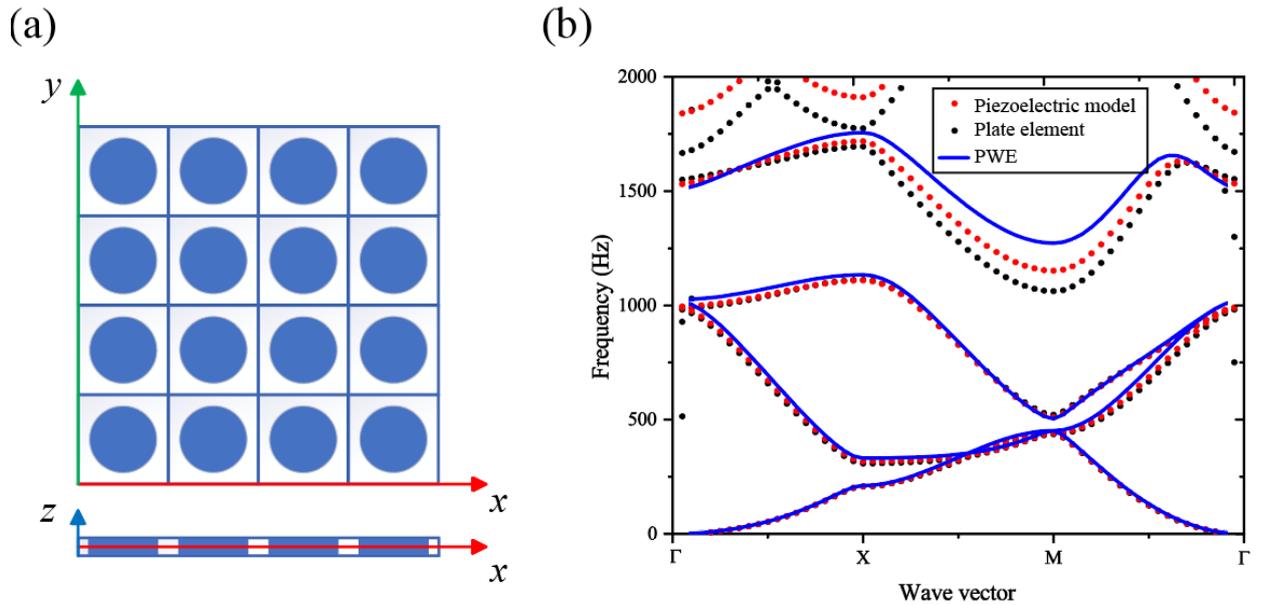

Fig. 3 Illustration of a two-phase phononic crystal plate (a). Band structures calculated by the 3D piezoelectric model (black dots), the 2D plate elements (red stars) and the PWE (blue line) method with simplifications (b).

Table 1 Material constants of the PZT-5H patches and the aluminum plate

| | PZT-5H | |
|---|---|---|
| **Constants** | **Values** | **Notions** |
| $s_{11}^E = s_{22}^E$, $s_{33}^E$, $s_{12}^E = s_{21}^E$, $s_{13}^E = s_{23}^E$, $s_{44}^E = s_{55}^E$, $s_{66}^E$ (m²/N) | $16.5 \times 10^{-12}$, $20.7 \times 10^{-12}$, $-4.78 \times 10^{-12}$, $-8.45 \times 10^{-12}$, $43.5 \times 10^{-12}$, $42.6 \times 10^{-12}$ | Compliance constants |
| $d_{31} = d_{32}$, $d_{33}$, $d_{24} = d_{15}$ (C/N) | $-274 \times 10^{-12}$, $593 \times 10^{-12}$, $741 \times 10^{-12}$ | Piezoelectric constants |
| $\varepsilon_{11}^T / \varepsilon_0 = \varepsilon_{22}^T / \varepsilon_0$, $\varepsilon_{33}^T / \varepsilon_0$ | 3130, 3400 | Dielectric permittivity constants |
| $\rho$ (kg/m³) | 7500 | Mass density |
| | **Aluminum** | |
| **Constants** | **Values** | **Notions** |
| $E$ (Pa) | $70 \times 10^9$ | Young's modulus |
| $\nu$ | 0.33 | Poisson's ratio |
| $\rho$ (kg/m³) | 2700 | Mass density |

To validate the results of the PWE method, numerical studies are also performed by using the commercial FEM package COMSOL Multiphysics. The 2D quadratic plate elements are used to solve Eq. (18). The piezoelectric model of the COMSOL Multiphysics using 3D or volume elements takes the electro-mechanical coupling into account. Therefore, the piezoelectric model is used to calculate the band structures of the MM plate as shown in Fig. 2 without simplifications. For the numerical simulation by COMSOL Multiphysics, the periodic boundary conditions based on the Bloch's theorem are applied to the boundaries of the unit-cell. Then, the eigenfrequency analysis by the MUMPS solver can be performed to obtain the band structures by sweeping all the edges of the first Brillion zone. In the following calculations, we choose a = 0.1m, $h_b = h_{pa} = h_{ps} = 0.001$m, and $r_{pa} = r_{ps} = 0.035$m. The material constants of the aluminum and PTZ-5H are given in Table 1.

The band structures of the square latticed MM plate calculated by the PWE method, the 2D quadratic Kirchhoff plate elements and the 3D quadratic piezoelectric model are shown in Fig. 3b. It can be seen that the results of the PWE method are not accurate compared to the results of the 3D piezoelectric model over the 4th band of the band structures. But a smaller difference can be found by comparing the results of the PWE method and the 2D plate elements. This means that the simplification of the uniform plate thickness is no longer applicable for high frequencies. In conclusion, under certain simplifications, the PWE method is fast and accurate to the 3rd band of the band structures.

To increase the accuracy for computing the band structures, we remove the above described simplifications step by step. Firstly, we remove the uniform thickness simplification by the FEM using the 2D plate elements with different plate thicknesses. Then, we remove the thin plate simplification and calculate the band structures by the FEM using the 3D solid elements. The dynamic equations of the wave motion in the 3D elastic solids are given by

$$\nabla \cdot \left[\mathbf{C}(\mathbf{r}) : \nabla \mathbf{u}(\mathbf{r},t)\right] = \rho(\mathbf{r}) \frac{\partial^2 \mathbf{u}(\mathbf{r},t)}{\partial t^2}, \tag{21}$$

where $\nabla = (\partial/\partial x, \partial/\partial y, \partial/\partial z)$ is the differential operator, $\mathbf{u}(\mathbf{r},t)$ is the displacement vector, $\mathbf{r} = (x, y, z)$ is the vector of Cartesian coordinates, $t$ is the time, $\mathbf{C}(\mathbf{r})$ and $\rho(\mathbf{r})$ are the elasticity tensor and mass density, respectively.

The band structures of the square-latticed and triangle-latticed MM plates calculated by the FEM are shown in Figs. 4(a) and 4(b). The black dots represent the results calculated by the 3D quadratic piezoelectric model. The blue and red dots are the results calculated by the FEM using the 3D quadratic solid and 2D quadratic plate elements, respectively. For the square lattice, the numerical results of the 3D quadratic solid and 2D quadratic plate elements are practically the same below the 4th band of the band structures. In comparison, the numerical results for the triangle lattice are also nearly the same below the 3rd band of the band structures. There is nearly no difference between the numerical results calculated by the 3D solid and 2D Kirchhoff plate elements below the 4th band for the square lattice or the 3rd band for the triangle lattice. This indicates that the uniformly distributed electric field in the PZT patches assumed in the simplified 2D plate model is sufficiently adequate in the low-frequency range. Therefore, the 3D solid elements can be used to calculate the band structures with complex geometry like the triangle lattice. The 2D Kirchhoff plate elements can be used when the computational efficiency is more important.

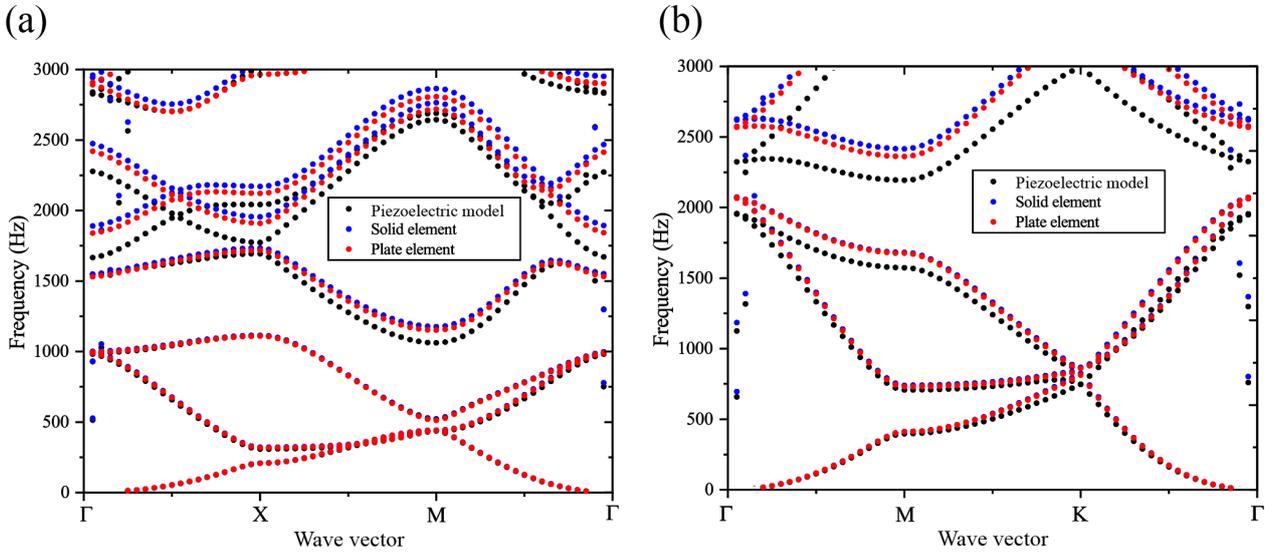

Fig. 4 Band structures of the square-latticed (a) and triangle-latticed (b) MM plates by the 3D piezoelectric model (black dots), the 3D solid elements (blue dots), and the 2D plate elements (red dots).

## 2.3 Wave transmission calculations

To verify the band structures and evaluate the vibration characteristics of the MM plate, we use the finite-size square-latticed MM plate structure as shown in Figs. 5(a) and 5(b) to calculate the wave transmissions or frequency responses in the $\Gamma X$ and $\Gamma M$ directions by using the 2D quadratic plate elements, respectively. To ensure the efficiency and accuracy, we choose 10 unit-cells for the wave transmission calculations. In the numerical simulation, the additional damping layers with the thickness $h_b$ and the artificial damping are introduced to eliminate the wave reflections by the finite boundaries of the phononic crystal structures [37].

The damping ratio is linearly increased from 0 to 5 outwards in the horizontal direction. The periodic boundary conditions are applied at the upper and lower boundaries in the y-direction. Because of the low-frequency limitation of our model which is caused by the uniform electric field simplification, the maximum frequencies 1691 Hz and 1626 Hz are used for the ΓX and ΓM directions, respectively. An out-of-plane acceleration is applied at the left boundary.

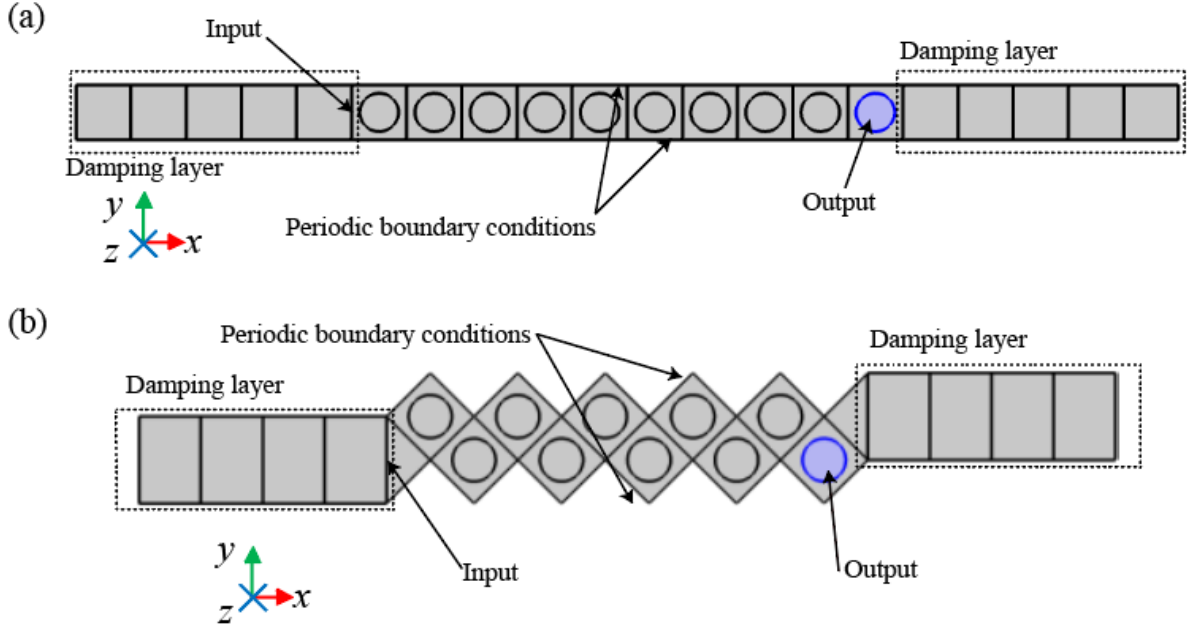

Fig. 5 Schematic diagrams of the transmission or frequency response calculations of the square-latticed MM plate in the ΓX direction (a) and ΓM direction (b). The domains in the dashed rectangles are the damping layers.

Then we define the transmission or normalized frequency response $f_{out}$ as

$$f_{out} = \log_{10} \frac{\iint_{S_{ps}} \frac{|a_z|}{\pi r_{ps}^2} dS}{a_{in}}, \tag{22}$$

where $S_{ps}$ is the area of the PZT sensor (shaded by blue color in Fig. 5), $a_z$ is the out-of-plane acceleration of the PZT sensor, and $a_{in}$ is the prescribed out-of-plane acceleration as input in Fig. 5. In our considered case, the input is taken as $a_{in}$ = 1 m/s².

Figure 6(a) illustrates the comparison between the transmission responses and the band structures. There are two directional band-gaps in the ΓX direction and one directional band-gap in the ΓM direction. It can be seen that 2nd band of the transmission response in the ΓX direction is wider than that of the band structure. This is because the 3rd band of the band structure, which is an anti-symmetric vibration mode in the *y*-direction as shown by the mode A in Fig. 6(b), cannot be excited by a uniform line source along the *y*-direction. Therefore, a deaf band is triggered [38]. Similarly, the displacement fields or vibration modes of the band edge states at the points B and C near the point M are shown in Fig. 6(b). It can be clearly seen that the mode B has the appropriate symmetry to be excited along the ΓM direction. In contrast, the mode C cannot be excited by the same uniform line source along the *y*-direction, and it appears again as a deaf band in the frequency transmission response. Because the deaf band in this case is very narrow and the attenuation is quite weak inside the deaf band in the ΓM direction, we will neglect this deaf band in the following analysis. Otherwise the transmission or frequency responses have a good agreement with the results of the computed band structures.

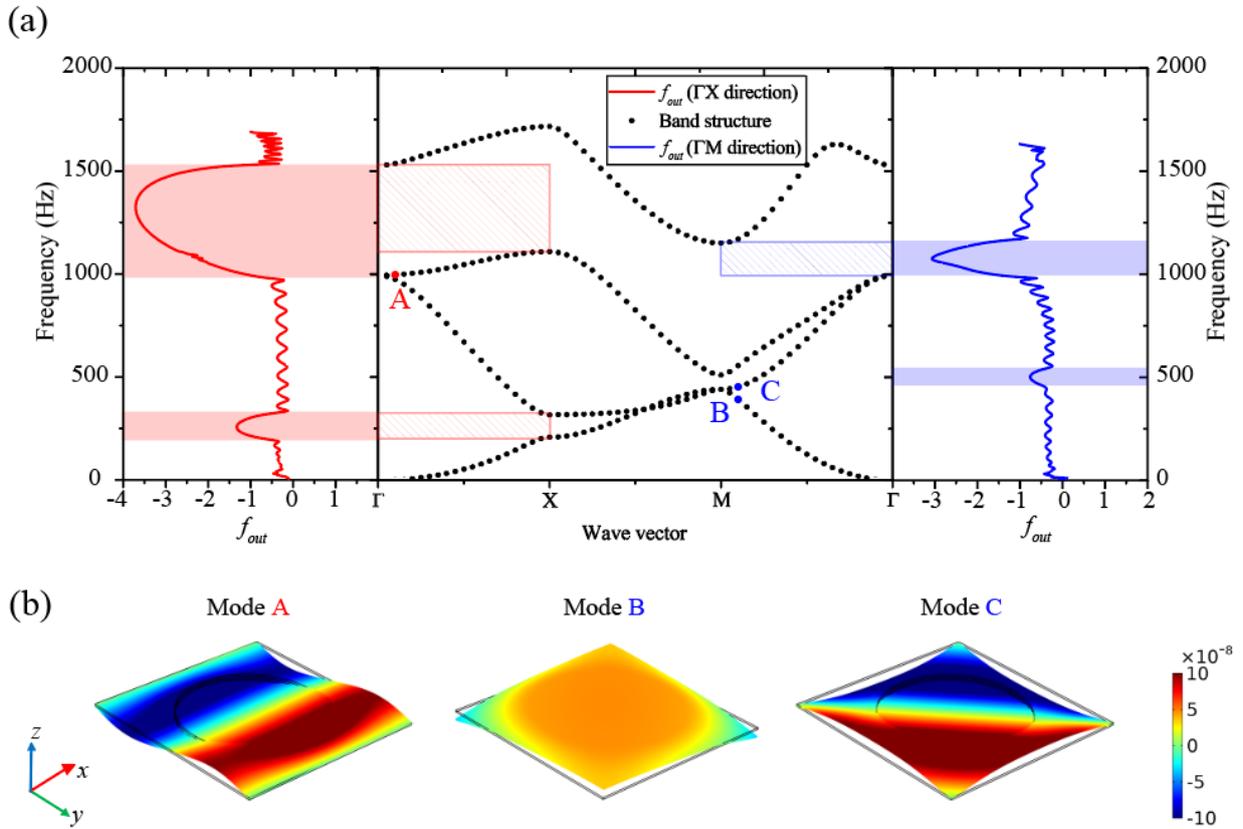

Fig. 6 Comparison between the transmission spectra or frequency responses and the band structures of the square-latticed MM plate (a). The red and blue lines are the results corresponding to the ΓX and ΓM directions, respectively. The band-gaps in the ΓX and ΓM directions are marked by red and blue, respectively. The displacement fields or vibration modes of the band edge states A, B and C in (a) are shown in (b).

In summary, a simplified model for the thin elastic plates covered by periodic piezoelectric patches based on the Kirchhoff's thin plate theory is developed. Further simplifications of the model include the uniform electric field and uniform height. The PWE and FEM are used to calculate the band structures. The FEM with 2D plate elements is used to calculate the wave transmission or frequency responses. By comparing with the 3D piezoelectric model, the PWE method is accurate in the low-frequency range up to the 3rd band of the band structures for the square lattice. The reason is that the simplifications of the uniform height and uniform electric field are no longer accurate for higher frequencies.

In comparison with the 3D piezoelectric model, the FEM with the 2D plate elements and 3D solid elements is accurate up to the 4th band for the square-latticed and 3rd band for the triangle-latticed MM plates, respectively. The FEM with the 2D plate elements and 3D solid elements shows nearly no differences in the computed band structures below the 4$^{th}$ band. Therefore, the FEM with the 2D plate elements is further used to calculate the band structures or transmission spectra in the following analysis. In the next section, we will introduce the active feedback control to the MM plate.

## 3. Active feedback control strategy

To enhance the dynamic performance of the MM plates, an active feedback control strategy is adopted in this analysis. Here, we apply a negative proportional feedback control strategy for the piezoelectric sensors and actuators. Accordingly, the $U_{zps}$ of the sensors is measured and applied to the piezoelectric actuators. Therefore, the $U_{zpa}$ of the actuators can be expressed as

$$U_{zpa} = -\left(g_d + g_v \frac{\partial}{\partial t} + g_a \frac{\partial^2}{\partial t^2}\right) U_{zps}, \quad (23)$$

where $g_d$ is the displacement feedback control gain, $g_v$ is the velocity or speed feedback control gain, and $g_a$ is the acceleration feedback control gain. Substituting Eqs. (16), (17) and (23) into Eq. (15), one can obtain the governing equation of motion for the MM plate as

$$\bar{m}_p \frac{\partial^2 w}{\partial t^2} + g_v \frac{\left(e_{31} C_{pa}^P\right)^2}{\varepsilon_{33}^S h_{pa}} \frac{\partial}{\partial t} \nabla^2 \nabla^2 w + \bar{D}_p \nabla^2 \nabla^2 w = 0, \qquad (24)$$

where

$$\bar{m}_p = m_p + g_a \frac{\left(e_{31} C_{pa}^P\right)^2}{\varepsilon_{33}^S h_{pa}} \nabla^2 \nabla^2, \qquad (25)$$

$$\bar{D}_p = D_b^E + D_{pa}^E + D_{ps}^E + \frac{\left(e_{31} C_{pa}^P\right)^2}{\varepsilon_{33}^S h_{pa}} (2 + g_d). \qquad (26)$$

It can be easily identified from Eqs. (24)-(26) that the displacement feedback control acts as an additional stiffness (last term in Eq. (26)), the velocity control introduces a damping related to the frequency (second term in Eq. (24)), and the acceleration control can be regarded as an additional mass related to the transverse displacement of the plate (second term in Eq. (25)).

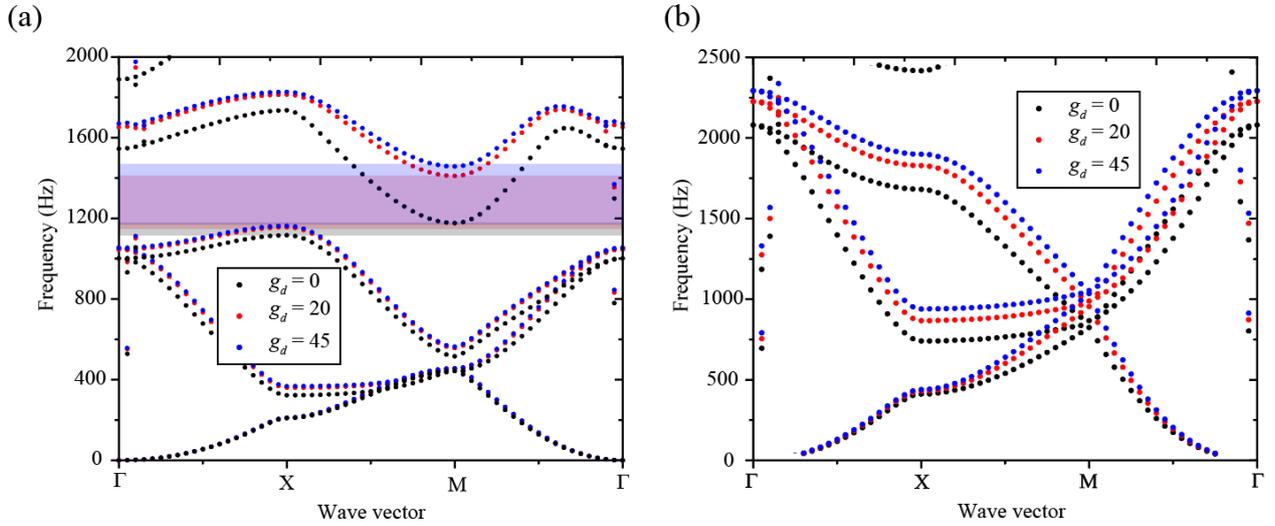

Fig. 7 Influences of $g_d$ on the band structures of the square-latticed (a) and triangle-latticed (b) MM plates. The shaded areas in (a) mark the band-gaps for $g_d = 0$ (gray), $g_d = 20$ (red) and $g_d = 45$ (blue).

The effects of $g_d$ on the band structures of the square-latticed and triangle-latticed MM plates are shown in Figs. 7(a) and 7(b), respectively. The bands shift to a higher frequency with the increase of $g_d$. The reason is that with the increase of $g_d$ the effective stiffness of the MM plate is enhanced. It can be seen from Fig. 7(a) that with the increase of $g_d$ the band-gap is widened and moved to a higher frequency range (1155.6-1310.2 Hz for $g_d = 0$ shaded with gray, and 1162.4-1457.3 Hz for $g_d = 45$ shaded with blue). The band-gaps in the $\Gamma X$ direction are changed from 211.3-344.4 Hz for $g_d = 0$ to 211.3-367.56 Hz for $g_d = 45$ and from 1155.6-1783.7 Hz for $g_d = 0$ to 1162.4-1826.9 Hz for $g_d = 45$. The band-gap in the $\Gamma M$ direction is changed from 1045.9-1310.2 Hz for $g_d = 0$ to 1054.3-1457.3 Hz for $g_d = 45$. There is no complete band-gap below the 3rd band of the band structures as shown in Fig. 7(b). The reason is that the stiffness increase doesn't break the asymmetry of the triangle lattice. But the band-gap in the $\Gamma M$ direction is changed from 431.2-741.5 Hz for $g_d = 0$ to 440.4-940.1 Hz for $g_d = 45$. As a common characteristic of all these band-gaps, the lower edge of the band-gaps is not sensitive to the change of $g_d$. Therefore, to achieve the widest band-gap the maximum admissible $g_d$ can be applied.

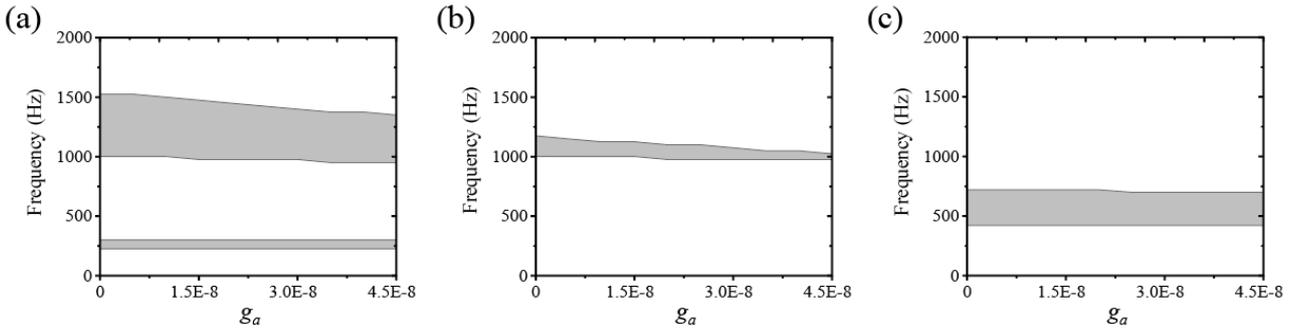

Fig. 8 The influences of $g_a$ on the band-gaps of the square lattice in the $\Gamma X$ direction (a), the square lattice in the $\Gamma M$ direction (b), and the triangle lattice in the $\Gamma M$ direction (c).

The effects of $g_a$ on the band-gaps are shown in Fig. 8. It can be seen from Fig. 8(a) that the first band-gap of the square lattice in the $\Gamma X$ direction is not sensitive to the change of $g_a$. In contrast, the second band-gap of the square lattice in the $\Gamma X$ direction shifts to a lower frequency as $g_a$ increases. The band-gap of the square lattice in the $\Gamma M$ direction as shown in Fig. 8(b) indicates the same changing trend as the band-gap in the $\Gamma X$ direction in Fig. 8(a). The band-gap of the triangle lattice in the $\Gamma M$ direction shown in Fig. 8(c) slightly shifts to a lower frequency as $g_a$ increases. This phenomenon is due to the fact that $g_a$ acts as an additional mass to the MM plate. With the increase of $g_a$, the effective mass is also increased. Therefore, the band-gap in Fig. 8(c) is shifted to a lower frequency. Since the increase of $g_a$ doesn't break the asymmetry of the triangle lattice, there is no band-gap in the ΓK direction. The band-gaps in the low-frequency range are less sensitive to the change of $g_a$. The active feedback control can only slightly affect the band structures of the triangle lattice. Therefore, the following calculations focus on the band structures of the square lattice.

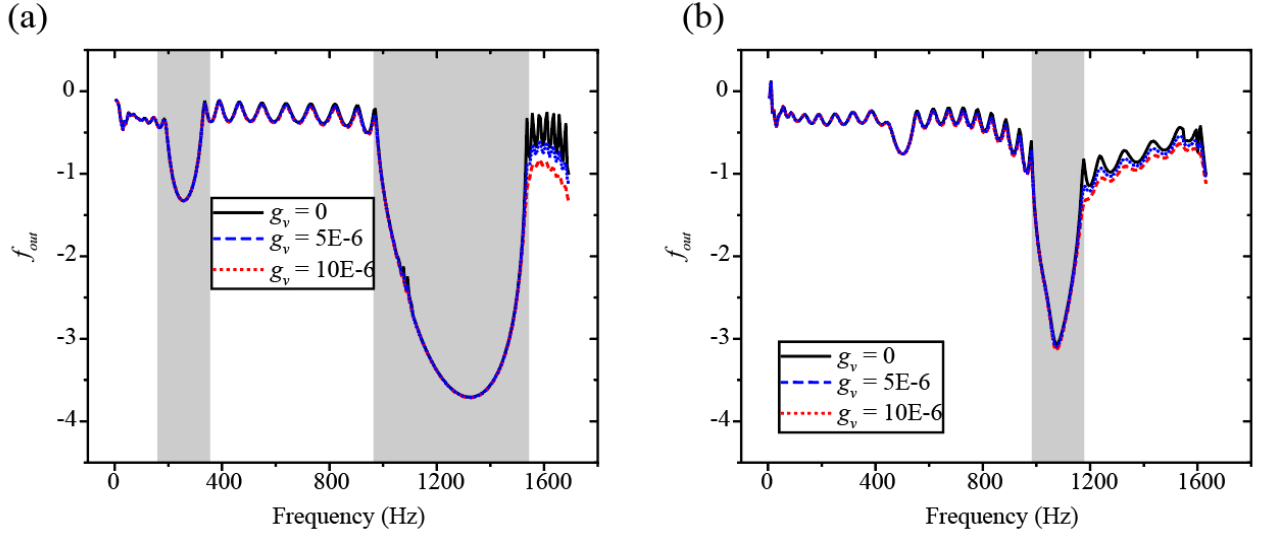

Fig. 9 Influences of $g_v$ on the $f_{out}$ of the square lattice in the ΓX direction (a) and the ΓM direction (b). The gray-shaded areas in (a) and (b) mark the band-gaps.

To show the effects of $g_v$ clearly, the frequency responses $f_{out}$ in the ΓX and ΓM directions are calculated and shown in Figs. 9(a) and 9(b), respectively. It can be seen that the $g_v$ does not affect the resonance peaks and band-gaps below 1500 Hz. And with the increase of $g_v$, the damping is also increased. It can be clearly identified from Eq. (24), that $g_v$ gives rise to an additional damping to the MM plate that linearly increases as the frequency increases.

To summarize, an active feedback control strategy for the MM plates is proposed in this section. The effects of the displacement, acceleration and displacement feedback control gains $g_d$, $g_a$ and $g_v$ are analyzed. Additional effective stiffness and mass are added as $g_d$ and $g_a$ are introduced, respectively. Therefore, the band-gaps shift to a higher frequency as $g_d$ increases, while the band-gaps shift to a lower frequency as $g_a$ increases. An additional damping is introduced into the MM plates as the velocity feedback control $g_v$ is the applied. Therefore, the best strategy to control the flexural waves in the MM plates is to apply the maximum admissible $g_d$ to achieve the widest band-gap at low frequencies and apply $g_v$ for higher frequencies. The displacement feedback control $g_v$ has very a weak influence on the band-gaps in the considered low-frequency range. Although the active feedback control strategy presented in this section can successfully suppress the flexural waves inside the band-gaps, each control method shows its own advantages and disadvantages. Besides, the flexural wave control by self-sensing and self-actuation is yet rarely reported in literature. For this reason, an active self-adaptive (ASA) control strategy is presented in the following section to suppress the flexural waves in the pass-bands outside of the band-gaps.

## 4. Active self-adaptive metamaterial

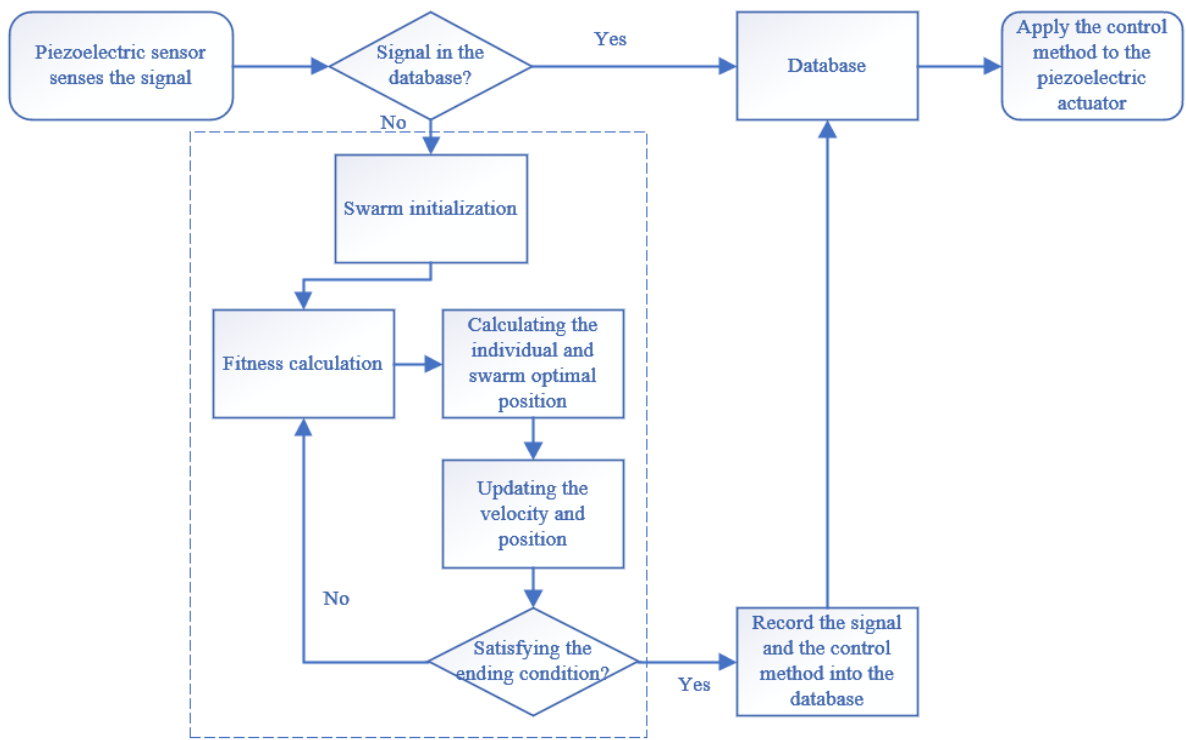

Fig. 10 Flow chart of the ASA control strategy under stimulation. The flow chart inside the dashed lines is the part of the PSO technique.

In the present active ASA control strategy, certain mechanisms should be introduced into the piezoelectric sensor and actuator to realize the self-sensing and self-actuation. Since the velocity feedback control gain $g_v$ has only very weak effects on the band structures as shown previously, we consider only the displacement feedback control gain $g_d$ and the acceleration feedback control gain $g_a$ in our ASA control strategy. To this end, we introduce the particle swarm optimization (PSO) technique for the piezoelectric sensor and actuator.

The goal of the PSO can be formulated as

$$\text{Minimize:} \quad f(g_d, g_a) = \sum_{f_{start}}^{f_{end}} f_{out},$$

$$\text{Subject to:} \quad g_d \in [0, 45], \quad g_a \in [0, 4.5\text{E}-8], \tag{27}$$

in which $f_{start}$ and $f_{end}$ are the start and end frequencies of the frequency range sensed by the piezoelectric sensor, respectively, and $\Sigma$ denotes the summation. Thus, the ASA control driven by the PSO technique is established. The flowchart of the ASA control strategy under stimulation is shown in Fig. 10. Based on the ASA control strategy, the ASAMM plates can be built. For a given stimulation, the ASA control strategy tries to find a corresponding control method from its database and then apply the control method to the piezoelectric actuator. If the stimulation is not contained the database, the PSO technique is used to generate a corresponding control method. Then the generated control method is applied to the actuator and recorded in the database.

The procedure of the PSO technique is shown inside the dashed lines in Fig. 10 and can be described by the following steps:

(i) An initial control gain is created randomly as the position and assigned to each particle. And a random number in the range of (0, max control gain) is created as the initial velocity for every particle.

(ii) The fitness of each particle is evaluated. Firstly, the frequency response $f_{out}$ is calculated by COMSOL Multiphysics. Then, the fitness is assigned by $f(g_d, g_a)$ defined in Eq. (27).

(iii) Find the smallest fitness for each particle, marked as the individual's optimal position (**pbest**). And find the smallest fitness for all particles, marked as the global optimal position (**gbest**).

(iv) Update the velocity and position as follows in the $i$th iteration

$$\mathbf{v}(i+1) = \omega \mathbf{v}(i) + c_p R_1(i) \left( \mathbf{pbest} - \begin{bmatrix} g_d(i) \\ g_a(i) \end{bmatrix} \right) + c_g R_2(i) \left( \mathbf{gbest} - \begin{bmatrix} g_d(i) \\ g_a(i) \end{bmatrix} \right), \tag{28}$$

$$\begin{bmatrix} g_d(i+1) \\ g_a(i+1) \end{bmatrix} = \begin{bmatrix} g_d(i) \\ g_a(i) \end{bmatrix} + \mathbf{v}(i+1), \tag{29}$$

in which $\omega$, $c_p$ and $c_g$ are parameters suggested by Ref. [39], $R_1$ and $R_2$ are independent and uniformly distributed random variables for each particle in the range of (0,1), $\mathbf{v}$ is the velocity vector for each particle which contains two velocity components of the corresponding control gains, and the hard boundary is used when the particles reach the maximum control gain [40].

(v) Repeat step (ii) to (iv) until an ending condition is satisfied. The ending condition requires that either all particles are in the same position or the maximum iteration number 100 is reached.

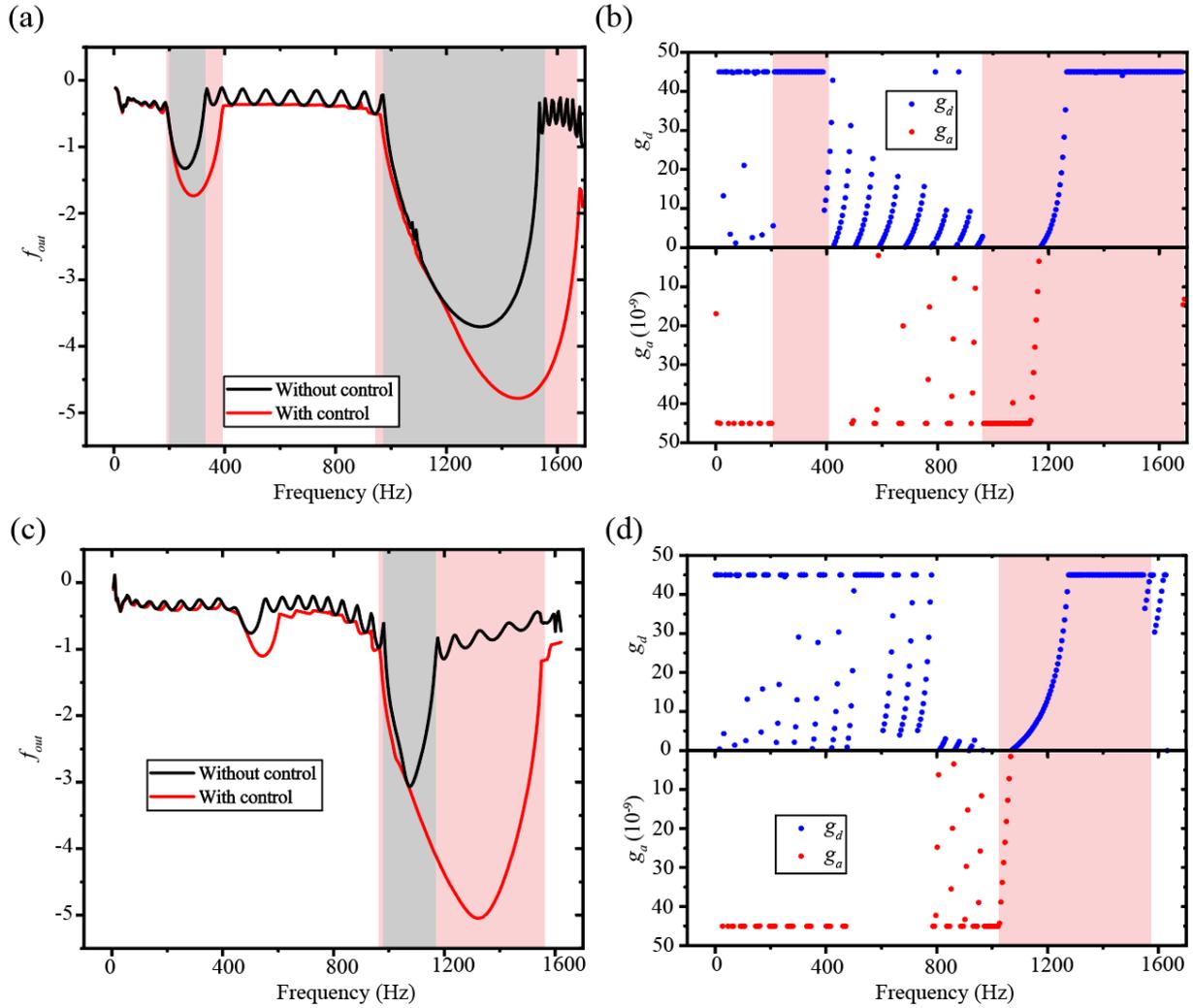

Fig. 11 Frequency responses in the ΓX direction (a) and ΓM direction (c) with damping layers. The control methods for (a) and (c) are presented in (b) and (d), respectively. The shaded areas in (a) and (c) designate the band-gaps without control (gray) and with control (red). The corresponding control methods are marked in red in (b) and (d).

For the situation that the ASAMM plate should work at a specific frequency, we chose $f_{str} = f_{end}$ in Eq. (27).

Therefore, all of the frequencies within the low-frequency limitation of our model are calculated. The transmission spectra or frequency responses $f_{out}$ in the ΓX and ΓM directions are shown in Figs. 11(a) and 11(c), where the damping layers are used as shown in Fig. 4(a). The corresponding control methods are shown in Figs. 11(b) and 11(d). It can be seen from Figs. 11(a) and 11(c) that the ASA control can largely widen the band-gaps. The damping inside the band-gaps is also enlarged remarkably.

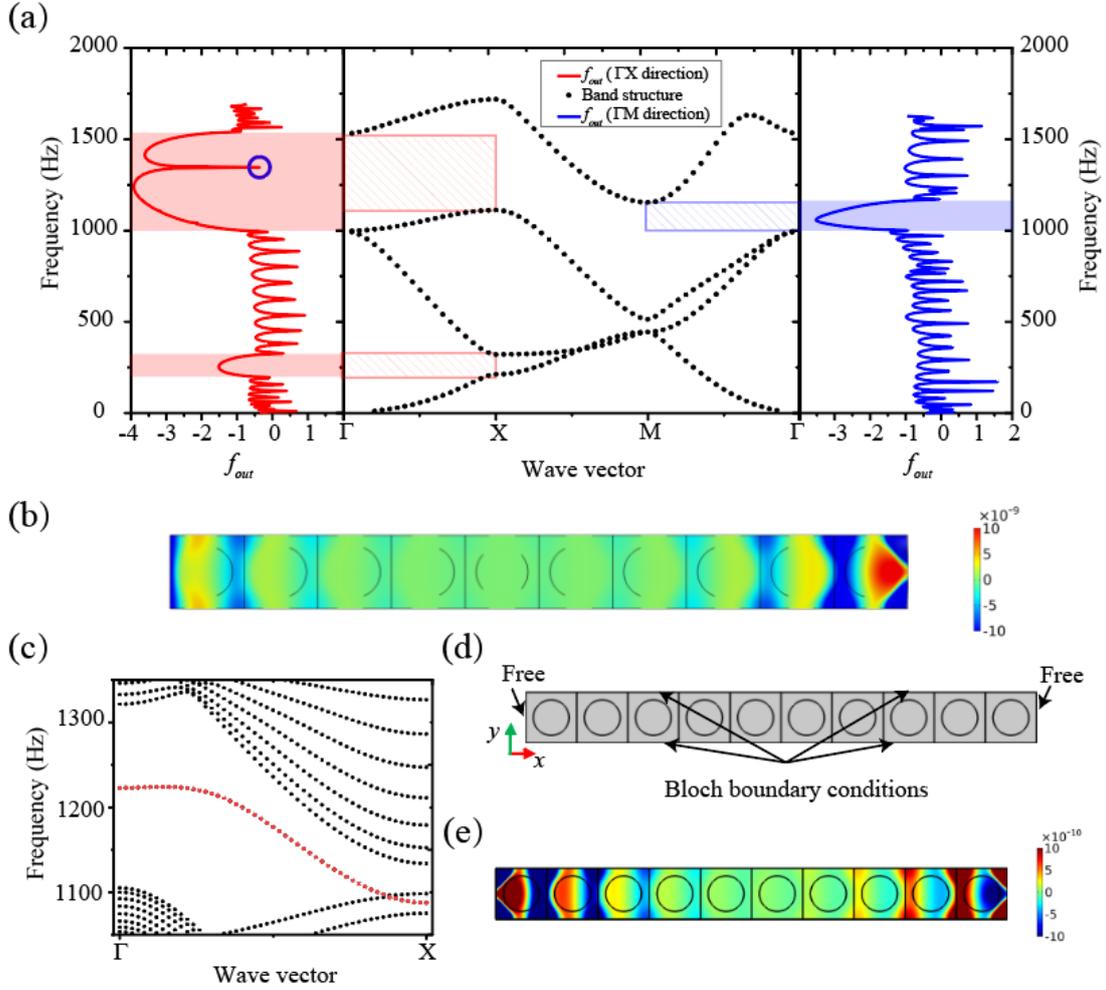

Fig. 12 Comparison between the frequency responses and the band structures without damping layers. The red line and blue line are the $f_{out}$ in the ΓX direction and ΓM direction, respectively. The band-gaps in the ΓX and ΓM directions are marked by red and blue. The displacement field at the frequency 1346 Hz of the peak, marked by the blue circle in (a), is shown in (b). The band structures of the unit-cell in the y-direction are shown in (c), while (d) illustrates the corresponding unit-cell with the free-free boundary conditions. The vibration mode of the red-doted line at the Γ point in (c) is shown in (e).

However, the resonance peaks of a finite phononic crystal structure cannot be predicted by the band structures because of the assumption of the infinite size in the band structure calculations. Therefore, the frequency response $f_{out}$ of the finite ASAMM plate without damping layers under free-free boundary conditions is also calculated and shown in Fig. 12(a). The $f_{out}$ without damping layers shows a good agreement with the computed band structures. However, a resonance peak in the ΓM direction is triggered as marked by the blue circle in Fig. 12(a), and its corresponding displacement field is shown in Fig. 12(b). We consider the finite structure with 10 unit-cells as shown in Fig. 12(d), and the band structures of the unit-cell are shown in Fig. 12(c). For the considered finite ASAMM plate, a new vibration mode is triggered and marked by the red-doted line. As shown in Fig. 12(e), the vibration is localized at the free boundaries of the finite ASAMM plate. It can be identified that this mode is an edge mode propagating in the $y$-direction. Moreover, for a finite ASAMM plate structure without damping layers, stronger resonance peaks can be found by comparing Fig. 12(a) with Fig. 6.

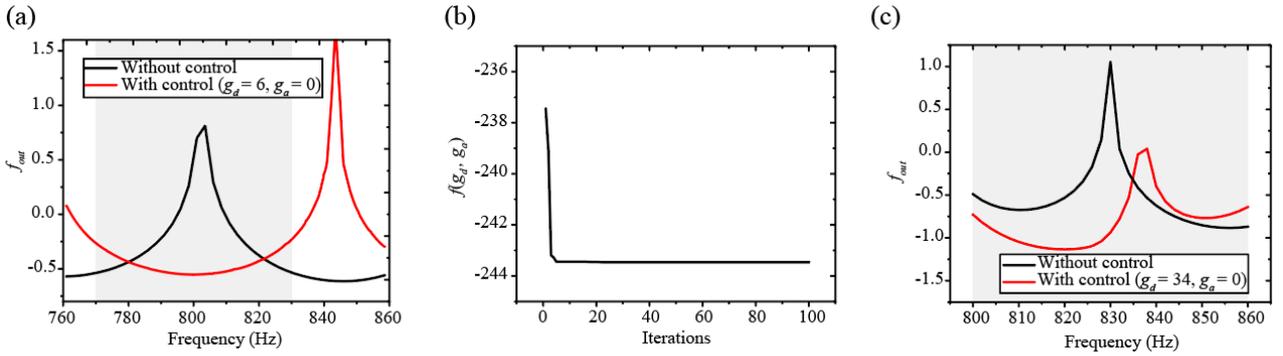

Fig. 13 Frequency responses of the ASAMM plate without damping layers in the ΓX direction in the frequency range 770-830 Hz shaded by gray (a) and in the ΓM direction in the frequency range 800-860 Hz (c). The convergence of the PSO technique for the ASAMM plate in the ΓX direction is shown in (b).

In the real-world applications, structural vibration is within certain frequency ranges, which may not lie in the band-gap. In this case, the vibration suppression at or near the resonance peaks is of particular importance. In the considered example, we choose a frequency range that contains a resonance peak in Fig. 12(a). Without loss of generality, the resonance peaks in the frequency range 770-830 Hz in the ΓX direction and 800-860 Hz in the ΓM direction are chosen. Both frequency ranges are outside of the band-gaps, which can be seen in Fig. 12(a). The transmission or frequency response results without and with the active self-adaptive control are shown in Figs. 13(a) and 13(c). It can be seen that the response peak is moved to a higher frequency by the control method ($g_d = 6$, $g_a = 0$) generated by the PSO technique in Fig. 13(b). Therefore, the vibration of the MM plate within the targeted frequency range is minimized. The convergence of the PSO technique to generate the control method for Fig. 13(a) is shown in Fig. 13(b). The PSO technique reaches the final result already at the 5th iteration. In most cases, the PSO technique can generate an acceptable control method within 10 iterations. As for the $\Gamma M$ direction, the PSO technique cannot move the resonance peak out of the targeted frequency range 800-860 Hz. Therefore, the control method with the smallest acceleration is chosen and applied ($g_d = 34$, $g_a = 0$) to reduce the resonance peak.

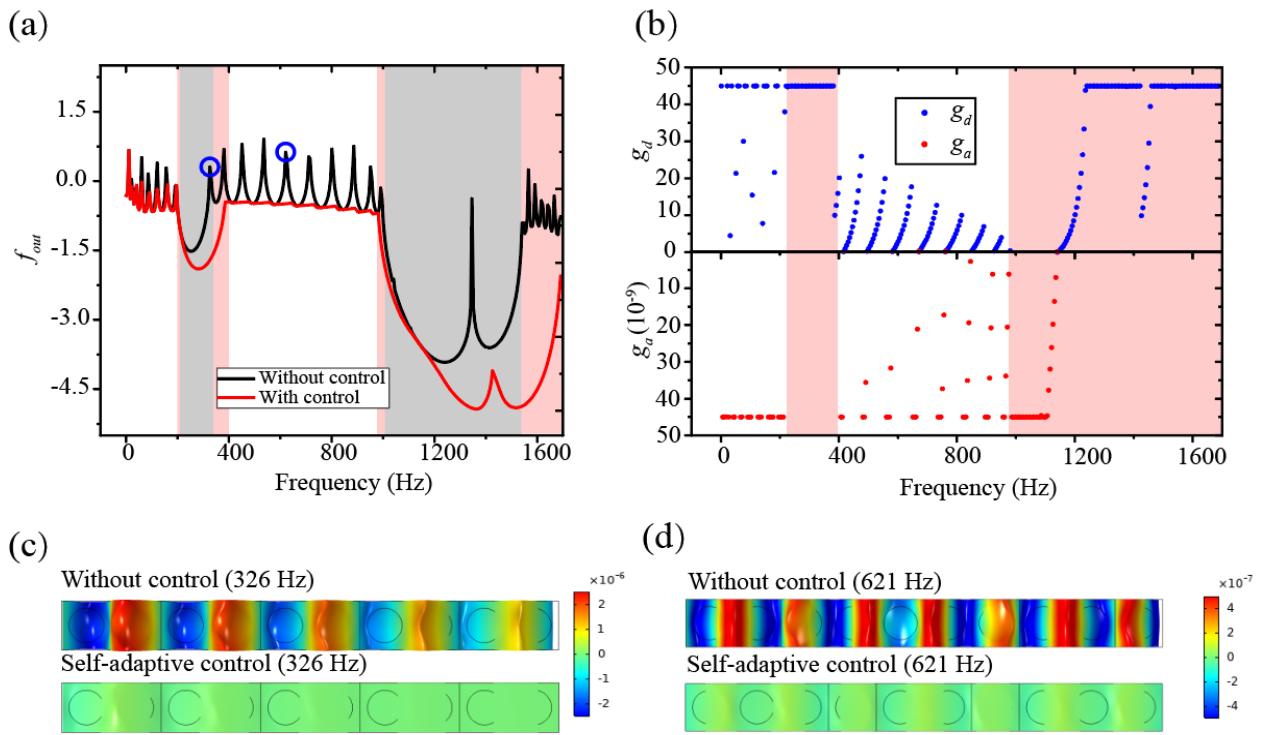

Fig. 14 Frequency responses of the ASAMM plate without damping layers in the $\Gamma X$ direction (a), and the corresponding control method (b). The displacement fields of the ASAMM plate at the resonance frequencies marked by the blue circles in (a) are shown in (c) and (d). The shaded areas in (a) mark the band-gaps without control (gray) and with control (red). The corresponding control methods for the band-gaps are marked in red in (b).

Next, the stimulation at a specific frequency is considered. Figure 14(a) illustrates the comparison of the frequency responses $f_{out}$ in the ΓX direction with and without the ASA control strategy. The corresponding control method is shown in Fig. 14(b). It can be seen from Fig. 14(a) that the ASA control strategy can widen the band-gap and enlarge the attenuation inside the band-gap. An interesting and important phenomenon is observed here, that the resonance peaks are practically eliminated, as shown in Fig. 14(a).

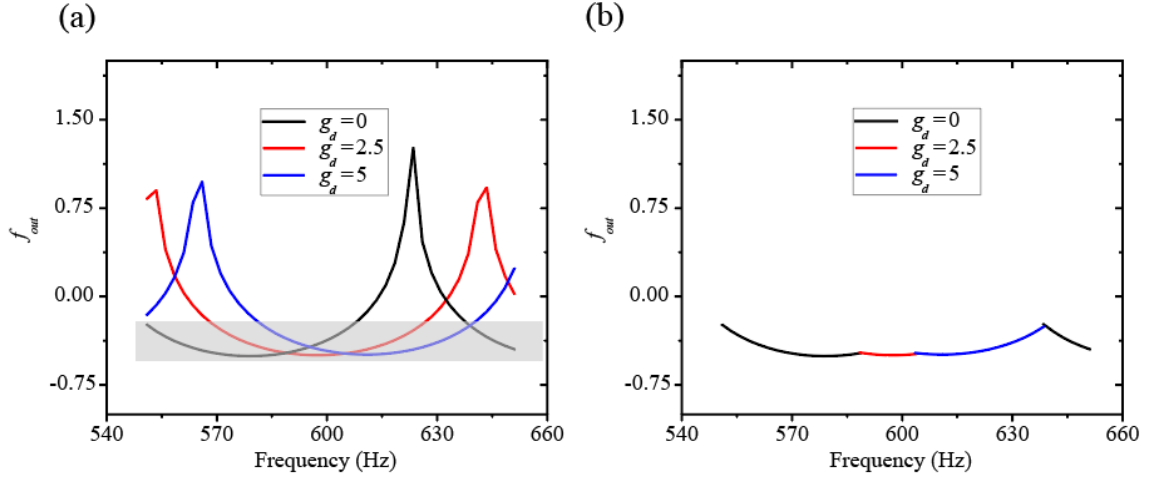

Fig. 15 The underlying physical mechanism for the elimination of the resonance peaks and lowest frequency response. The frequency responses $f_{out}$ for different control gains (a). The frequency response $f_{out}$ with assigned control gains by ASA control (b).

There are two mechanisms for the drastic reduction or nearly complete elimination of the resonance peaks. The first one is that the ASA control strategy widens the band-gap. At 326 Hz as marked by the blue circle in Fig. 14(a), this frequency falls into the new band-gap as the control method is applied. Therefore, the resonance peak at 326 Hz disappears, as illustrated by the displacement field in Fig. 14(c). The second one is that the resonance peak is moved to another frequency. This is because that the active feedback control changes the effective stiffness or the effective density of the finite structure. Consequently, the positions of the resonance peaks are also altered as shown in Fig. 15(a). The ASA control strategy can smartly choose the appropriate control gain leading to the lowest frequency response $f_{out}$ as shown in the gray area in Fig. 15(a). Then, the ASA control strategy will assign these chosen appropriate control gains to different frequencies as shown in Fig. 15(b). Finally, the resonance peaks are eliminated and the lowest frequency response $f_{out}$ in the considered frequency range is achieved. Therefore, the resonance peak in Figure 14(d) is largely reduced.

Moreover, the edge mode at 1346 Hz in Fig. 12(a) is also eliminated. However, the resonance peaks at extremely low frequencies (less than 200 Hz in our considered case) cannot be completely eliminated by the proposed ASA control strategy. The reason is that the proposed active feedback control has only a very slight effect on the flexural wave propagation at extremely low frequencies, as shown in Figs. 7(a) and 8(a).

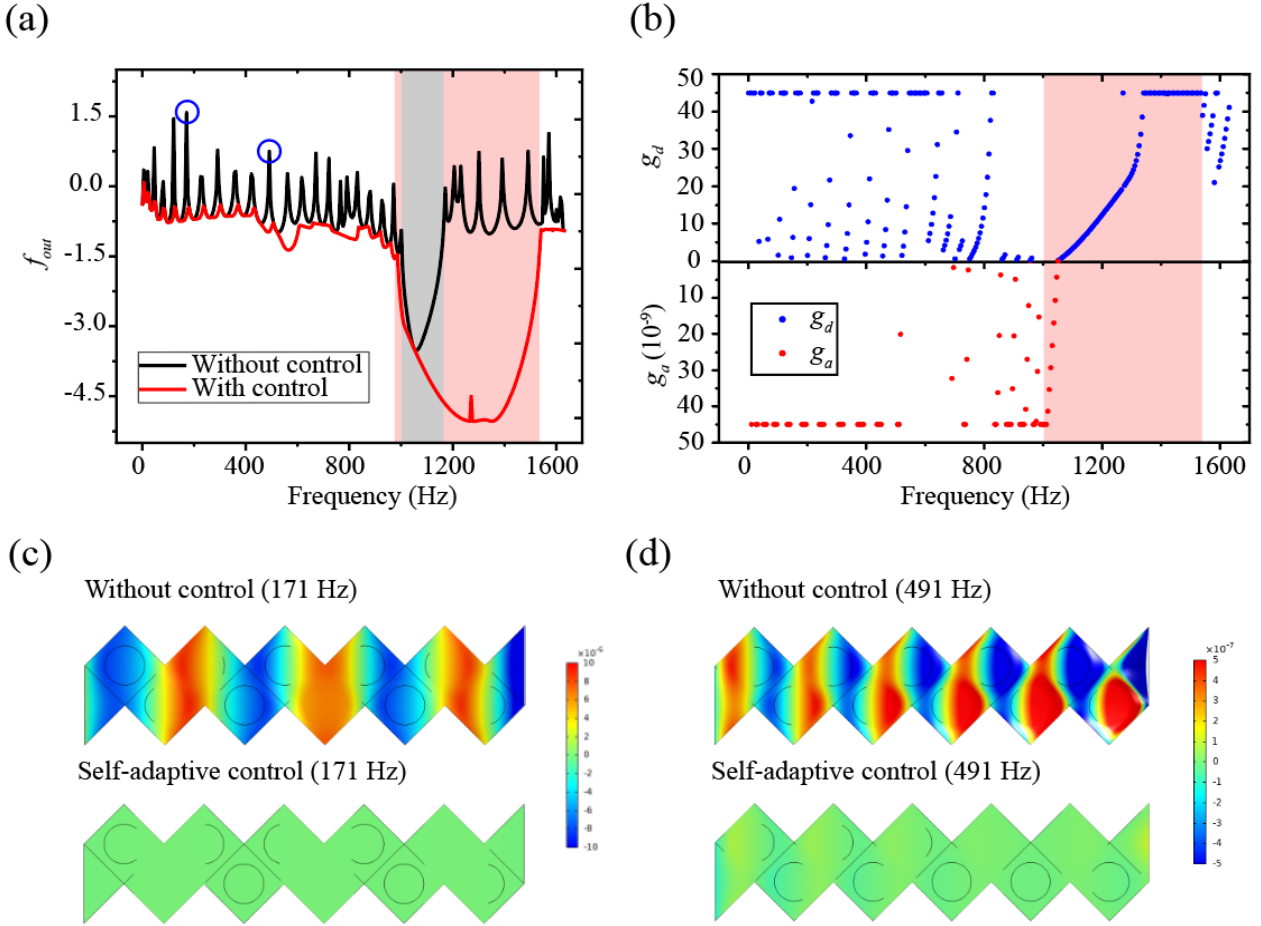

Fig. 16 Frequency responses of the ASAMM plate in the ΓM direction (a), and the corresponding control method (b). The displacement fields of the ASAMM plate at the resonance frequencies marked by the blue circles in (a) are shown in (c) and (d). The shaded areas in (a) mark the band-gaps without control (gray) and with control (red). The corresponding control methods for the band-gaps are marked in red in (b).

Figure 16(a) illustrates the comparison of the frequency responses $f_{out}$ in the ΓM direction with and without the ASA control strategy, and the corresponding control method is given in Fig. 16(b). The ASA control strategy can substantially widen the band-gap and enlarge the attenuation inside the band-gap. The resonance peaks are also drastically diminished or nearly eliminated. Figures 16(c) and 16(d) show the displacement fields at 171 and 491 Hz respectively. It can be seen clearly that the vibration is restrained.

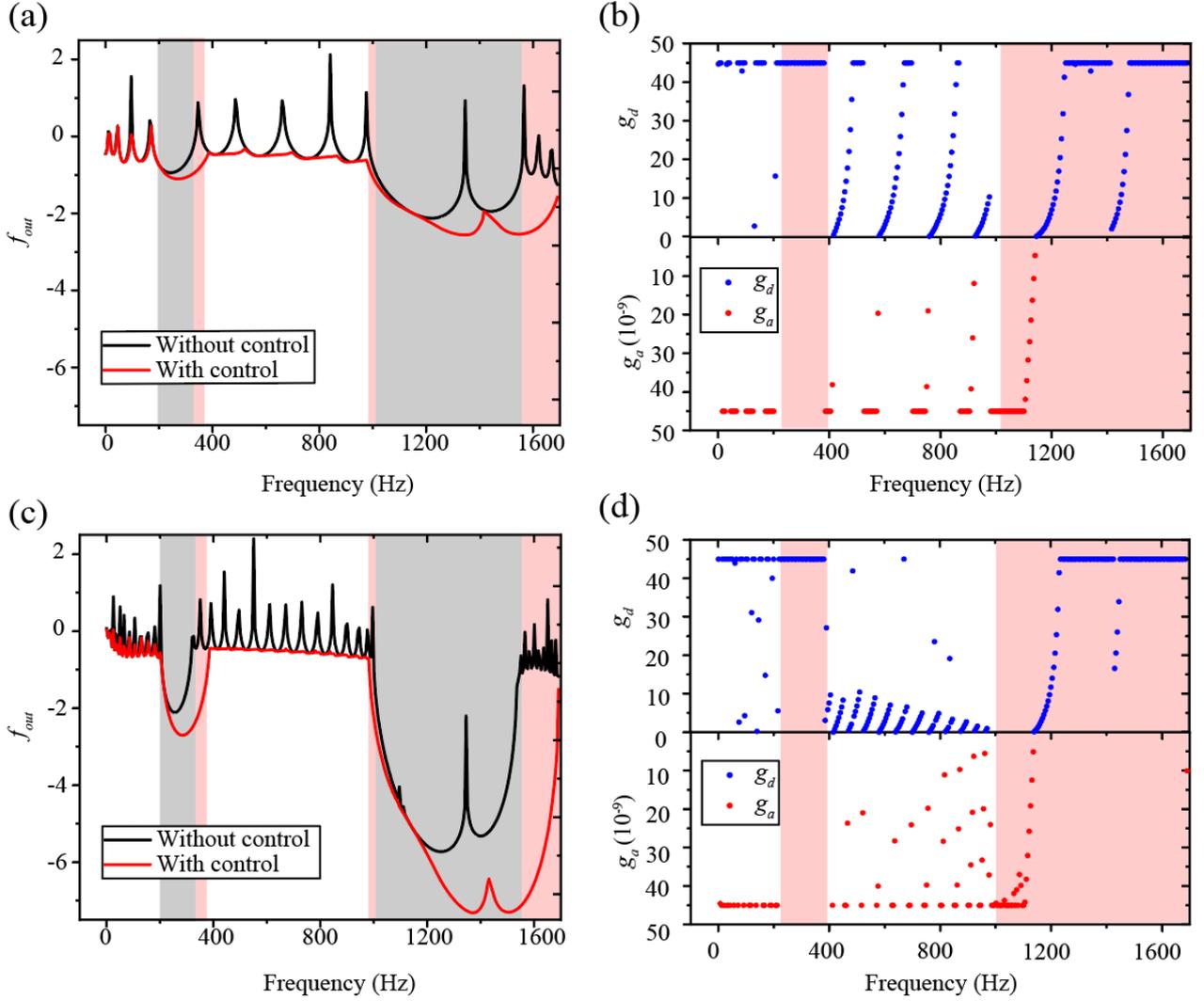

Fig. 17 Frequency responses of the ASAMM plate in the ΓX direction with 5 unit-cells (a) and 15 unit-cells (c) without damping layers. The corresponding control methods for (a) and (c) are presented in (b) and (d), respectively. The shaded areas in (a) and (c) mark the band-gaps without control (gray) and with control (red). The corresponding control methods for the band gaps are marked in red in (b) and (d).

By comparing Figs. 6(a) and 7(a) with Figs. 14(a) and 15(a), one can find that the edges of the enlarged band-gaps are determined by the displacement and acceleration feedback control gains $g_d$ and $g_a$. In our considered cases, $g_d$ mainly determines the upper edge while $g_a$ dominantly determines the lower edge of the band-gaps. This indicates that the proposed ASA control strategy cannot generate new band-gaps but can widen the band-gaps without control. Inside the band-gap, the proposed ASA strategy can automatically choose the appropriate control method with the largest attenuation.

Finally, we analyze the effect of the number of the unit-cells in the finite ASAMM plate by calculating and comparing the transmission spectra or frequency responses $f_{out}$ for 5 and 15 unit-cells with that for 10 unit-cells as presented previously. Figure 16 illustrates the frequency responses $f_{out}$ of the ASAMM plates with 5 unit-cells (Fig. 17(a)) and 15 unit-cells (Fig. 17(c)) in the $\Gamma X$ direction. By comparing Figs. 17(a) and 17(c) with Fig. 14(a) we can conclude that the location and the width of the first two band-gaps are nearly unaffected by the number of the unit-cells constituting the ASAMM plate, and 10 unit-cells are sufficient for calculating and identifying the first two band-gaps. This justifies the adequacy of the previously analyzed finite ASAMM plate structures with 10 unit-cells for the band-gap analysis. In contrast, the number and the location of the resonance peaks are strongly dependent on the number of the unit-cells used in the transmission or frequency response calculations. One can find that the proposed ASA control strategy can reduce or nearly eliminate the resonance peaks for the cases with 5 and 15 unit-cells. Figures 17(b) and 17(d) show the corresponding control methods for Figs. 17(a) and 17(c). The resonance peak related to the edge mode at the frequency 1346 Hz is also reduced and shifted. However, the resonance peaks inside or outside of the band-gaps can be significantly reduced or nearly eliminated in all three considered cases with 5, 10 and 15 unit-cells by using the proposed ASA control strategy automatically, independent of the number of the unit-cells used in the ASAMM plates.

In summary, this section presents an efficient design method for the ASAMM plates based on the ASA control strategy driven by the PSO technique. The ASAMM can automatically evolve different control methods combining the displacement and acceleration feedback control schemes to match specific requirements without supervision, such as particularly targeted frequency ranges and different numbers of unit-cells etc. The designed ASAMM plates have three essential advantages comparing to the traditional MM plates. Firstly, the proposed ASA control strategy expands the band-gaps of the ASAMM plates. Secondly, the wave attenuation inside the band-gap is increased. Thirdly, the ASA control strategy suppresses the vibration outside the band-gaps by drastically reducing or nearly eliminating the resonance peaks.

## 5. Conclusions

In this paper, active self-adaptive metamaterial (ASAMM) plates are proposed and investigated, which consist of a thin base plate and two periodic arrays of piezoelectric patches attached on the top and bottom plate surfaces. The periodic piezoelectric patches on the bottom plate surface act as sensors, while the other ones on the top plate surface act as actuators. A simplified model is first developed to calculate the band structures and frequency responses of the ASAMM plates. The PWE method and the FEM are used for the numerical calculations. Based on the simplification of a uniform electric field in the piezoelectric patches, the FEM using 3D solid elements is accurate to the 4$^{th}$ band of the band structures. Furthermore, based on the Kirchhoff's thin plate theory, the FEM using 2D plate elements provides practically the same results as the FEM using 3D solid elements. Since the PWE method cannot handle the thickness discontinuities, a uniform plate thickness is further assumed based on the previous simplifications. The results show that the PWE method is accurate to the 3$^{rd}$ band of the band structures.

Then, based on the FEM using quadra elements, the active feedback control of the MM plates is analyzed. Three active feedback control methods, namely the displacement, acceleration and velocity feedback control methods with the feedback control gains $g_d$, $g_v$ and $g_a$ are introduced and investigated. An additional effective stiffness, an additional effective mass and an additional effective damping are induced as $g_d$, $g_a$ and $g_v$ are applied, respectively. Consequently, the band-gaps shift to a higher frequency as $g_d$ increases, while to a lower frequency as $g_a$ increases. In the low-frequency range, the effects of the velocity feedback control $g_v$ on the band-gaps and frequency responses are negligibly small.

Then, an active self-adaptive (ASA) control strategy for the ASAMM plates is developed based on the PSO technique. The ASA control strategy can automatically generate different control methods by combining the displacement and acceleration feedback control schemes to match different situations and requirements. The designed ASAMM plates based on the proposed ASA control strategy have three advantages compared with the conventional MM plates based on the separate active feedback control methods. Firstly, the band-gaps can be widened, and the edges of the widened band-gaps are determined by the feedback control gains $g_d$ and $g_a$. Secondly, the attenuation inside the band-gap is increased. Thirdly, the vibration outside the band-gaps is minimized by eliminating the resonance peaks of the finite ASAMM plate structures.

The simplified MM plate model based on the active feedback control methods can be used as a benchmark for further related and advanced studies. The proposed ASA control strategy can be extended and applied to design novel ASA metasurface, lenses, and other acoustic/elastic wave devices.

**Acknowledgements**

The present work was supported by the German Research Foundation (DFG, Project-No. ZH 15/30-1), and the National Natural Science Foundation of China (Project-No. 11761131006). T.X. Ma is also grateful to the support by German Research Foundation (DFG, Project-No. ZH 15/27-1).